\documentclass[11pt,a4paper,english,showpacs,superscriptaddress,aps]{revtex4}

\usepackage{amssymb}
\usepackage{amsmath}
\usepackage{graphicx}
\usepackage{babel}
\usepackage{hyperref}

\begin{document}

\title{Coleman-Weinberg mechanism in a three-dimensional supersymmetric Chern-Simons-matter model}

\author{A.~F.~Ferrari}
\email{alysson.ferrari@ufabc.edu.br}
\affiliation{Centro de Ci\^encias Naturais e Humanas, Universidade Federal do ABC, Rua Santa Ad\'elia, 166, 09210-170, Santo Andr\'e, SP, Brazil}

\author{E.~A.~Gallegos}
\email{gallegos@fma.if.usp.br}
\affiliation{Instituto de F\'{\i}sica, Universidade de S\~{a}o Paulo\\
Caixa Postal 66318, 05315-970, S\~{a}o Paulo - SP, Brazil}

\author{M.~Gomes}
\email{mgomes@fma.if.usp.br}
\affiliation{Instituto de F\'{\i}sica, Universidade de S\~{a}o Paulo\\
 Caixa Postal 66318, 05315-970, S\~{a}o Paulo - SP, Brazil}

\author{A.~C.~Lehum}
\email{andrelehum@ect.ufrn.br}
\affiliation{Escola de Ci\^encias e Tecnologia, Universidade Federal do Rio Grande do Norte\\
Caixa Postal 1524, 59072-970, Natal, RN, Brazil}

\author{J.~R.~Nascimento} 
\email{jroberto@fisica.ufpb.br}
\affiliation{Departamento de F\'{\i}sica, Universidade Federal da Para\'{\i}ba\\
 Caixa Postal 5008, 58051-970, Jo\~ao Pessoa, Para\'{\i}ba, Brazil}

\author{A. Yu. Petrov}
\email{petrov@fisica.ufpb.br}
\affiliation{Departamento de F\'{\i}sica, Universidade Federal da Para\'{\i}ba\\
 Caixa Postal 5008, 58051-970, Jo\~ao Pessoa, Para\'{\i}ba, Brazil}

\author{A.~J.~da~Silva}
\email{ajsilva@fma.if.usp.br}
\affiliation{Instituto de F\'{\i}sica, Universidade de S\~{a}o Paulo\\
 Caixa Postal 66318, 05315-970, S\~{a}o Paulo - SP, Brazil}

%\date{\today}

\begin{abstract}
Using the superfield formalism, we study the dynamical breaking of gauge symmetry {and superconformal invariance} in the ${\cal N}=1$ three-dimensional supersymmetric Chern-Simons model, coupled to a complex scalar superfield with a quartic self-coupling. This is an analogue of the conformally invariant Coleman-Weinberg model in four spacetime dimensions. We show that a mass for the gauge and matter superfields are dynamically generated after two-loop corrections to the effective superpotential. We also discuss the ${\cal N}=2$ extension of our work, showing that the Coleman-Weinberg mechanism in such model is not feasible, because it is incompatible with perturbation theory.
\end{abstract}


\maketitle

\section{Introduction}

The mechanism of spontaneous symmetry breaking is essential to explain the origin of masses of fermions and vector bosons in the standard model, and it is the reason why the Higgs particle, that is hoped to be discovered by the LHC in the next years, was postulated. In 1973~\cite{Coleman:1973jx}, Coleman and Weinberg (CW) discussed the very appealing scenario that radiative corrections could naturally induce this symmetry breaking. Their discussion was focused on four-dimensional models, but the mechanism of dynamical generation of mass in three-dimensional models was also considered afterwards~\cite{Tan:1996kz,Tan:1997ew,Dias:2003pw,Alves:1999hw,deAlbuquerque:2000ec}. 

Three-dimensional fields models are interesting for offering a simpler setting to study properties of gauge theories, including the possibility of the introduction of a topological mass (Chern-Simons) term~\cite{Deser:1981wh,Schonfeld:1981} in the Lagrangian. From a more practical viewpoint, Chern-Simons theories in three-dimensions were applied to the understanding of the quantized Hall effect~\cite{prange}. More recently, supersymmetric Chern-Simons models~\cite{Schwarz:2004yj,Bandres:2008vf} have been on focus due to its duality with gravity~\cite{Aharony:2008ug}. Conformal Chern-Simons theory have been explored in the construction of a theory modeling M2-branes~\cite{Bagger:2006sk,Bagger:2007jr,Gustavsson:2007vu}, and Chern-Simons gravities were coupled to 2p-branes~\cite{Miskovic:2009dd}. When studying some three-dimensional gauge theories coupled to a scalar field, in such a way that no dimensionful parameters appear in the classical Lagrangian, it was found that quantum corrections dynamically introduce a mass for the vector and scalar particles~\cite{Tan:1996kz,Tan:1997ew,Dias:2003pw,Alves:1999hw,deAlbuquerque:2000ec}. Differently from what happens in four-dimensional models, where the gauge symmetry is dynamically broken already at one-loop order, in three dimensions this mechanism appears only when two-loop corrections to the effective potential are taken into account.

It was long ago shown that the three-dimensional massless supersymmetric quantum electrodynamics and Wess-Zumino models do not exhibit dynamical generation of mass up to one-loop level~\cite{Burgess:1983nu}. Recently, in~\cite{Lehum:2008vn}, one of us examined the two-loop quantum corrections to the $D=(2+1)$ Wess-Zumino model and found the existence of dynamical generation of mass, via Coleman-Weinberg mechanism, and in~\cite{Lehum:2008ue} it was shown that the massive three-dimensional supersymmetric quantum electrodynamics exhibit a spontaneous gauge symmetry broken phase. This motivates the investigation of whether dynamical generation of mass also happens in higher loop levels in the three-dimensional supersymmetric Chern-Simons-matter model (SCSM).

In this work, we will study the possibility of dynamical breaking of the gauge symmetry and of supersymmetry in a three-dimensional supersymmetric Chern-Simons model coupled to a complex scalar matter superfield with a quartic self-interaction. This is the simplest supersymmetric three-dimensional analog to the conformally invariant model studied by Coleman-Weinberg~\cite{Coleman:1973jx}, where no mass scale appears in the classical Lagrangian. Our results are that the superconformal and gauge symmetries admits a broken phase but supersymmetry does not. We also show that the CW mechanism of breakdown of the gauge and superconformal symmetries does not work for the ${\cal{N}}=2$ extension of the SCSM, in agreement with the results obtained by Gaiotto and Yin~\cite{Gaiotto:2007qi} {and by Buchbinder et al.~\cite{Buchbinder:2010em}.} 

The paper is organized as follows. In Sec. \ref{modeldefs} we will define and study the ${\cal{N}}=1$ SCSM model. In Sec. \ref{N=2} we will deal with the ${\cal{N}}=2$ extension of SCSM model~\cite{Lee:1990it,Gates:1991qn,Nishino:1991sr}, and in Sec. \ref{final} we present a discussion and our final remarks.

\section{The supersymmetric D=(2+1) Chern-Simons-matter model}\label{modeldefs}

Our starting point is the classical action
\begin{eqnarray}\label{ceq1}
S&=&\int{d^5z}\Big{\{}A^{\alpha}W_{\alpha}
-\frac{1}{2}\overline{\nabla^{\alpha}\Phi}\nabla_{\alpha}\Phi
+\lambda(\bar\Phi\Phi)^2\Big{\}},
\end{eqnarray}

\noindent
where  $W^{\alpha}=(1/2)D^{\beta}D^{\alpha}A_{\beta}$ is the gauge superfield strength and $\nabla^{\alpha}=(D^{\alpha}-ieA^{\alpha})$ it is the supercovariant derivative. As we are not dealing with topological aspects of the model, we absorbed the Chern-Simons level parameter $\kappa$ into the dimensionless coupling $e=e'/\sqrt{\kappa}$ considered to be small. We use the notations and conventions contained in~\cite{Gates:1983nr}. The component expansions of the superfields involved in our work are provided in appendix~\ref{sup}. 
This action possesses manifest ${\cal{N}} = 1$ supersymmetry, but this symmetry can be lifted to ${\cal{N}}=2$ by the elimination of the fermion-number violating terms~\cite{Lee:1990it}, identifying the coupling constants according to $\lambda=-e^2/8$. This extension will be treated in the next section.

The model defined in Eq.(\ref{ceq1}) is invariant under the following infinitesimal $U(1)$ gauge transformations,
\begin{eqnarray}\label{ceq2}
\bar\Phi\longrightarrow \bar\Phi^{\prime}=\bar\Phi(1-ie \Lambda)~,\nonumber\\
\Phi\longrightarrow \Phi^{\prime}=(1+ie \Lambda)\Phi~,\\
A_{\alpha}\longrightarrow A_{\alpha}^{\prime}=A_{\alpha}+D_{\alpha}\Lambda~,\nonumber
\end{eqnarray}

\noindent
where the gauge parameter $\Lambda=\Lambda(x,\theta)$ is a real scalar superfield.

At tree level the gauge symmetry of this model is not spontaneously broken, differently from what happens when a mass term is included in the classical action~\cite{Lehum:2007nf}. Therefore we will investigate if it can be broken by radiative corrections. {To this end, we shift the superfields $\bar\Phi$ and $\Phi$ by the classical background field}
\begin{eqnarray}
\label{sigmacl}
\sigma_{cl}=\sigma_1-\theta^2\sigma_2~,
\end{eqnarray}

\noindent
where $\sigma_1$ and $\sigma_2$ are real constants. Rewriting them in terms of two real quantum fields $\Sigma$ and $\Pi$ in the form,
\begin{eqnarray}\label{ceq5}
\bar\Phi&=&\frac{1}{\sqrt{2}}\Big(\Sigma+\sigma_{cl}-i\Pi\Big)\nonumber\\
\Phi&=&\frac{1}{\sqrt{2}}\Big(\Sigma+\sigma_{cl}+i\Pi\Big)~,
\end{eqnarray}

\noindent
we assume the vanishing of the vacuum expectation values (VEV) of the quantum superfields, i.e., $\langle\Sigma\rangle=\langle\Pi\rangle=0$ at any order of perturbation theory. In terms of $\Sigma$ 
and $\Pi$ an infinitesimal gauge transformation results in
\begin{eqnarray}\label{ceq5a}
\Pi &\longrightarrow & \Pi^{\prime}=\Pi+e\Lambda(\Sigma+\sigma_{cl})~,\nonumber\\
\Sigma &\longrightarrow & \Sigma^{\prime}=\Sigma-e\Lambda\Pi~,
\end{eqnarray}

\noindent
and the gauge invariant action (\ref{ceq1}) results in,
\begin{eqnarray}\label{ceq6}
S&=&\int{d^5z}\Big{\{}A^{\alpha}W_{\alpha}-\frac{e^2\sigma_{cl}^2}{4}A^{\alpha}A_{\alpha}
-\frac{e\sigma_{cl}}{2}D^{\alpha}A_{\alpha}\Pi
+\frac{1}{2}\Sigma(D^2+3\lambda\sigma_{cl}^2)\Sigma+\frac{1}{2}\Pi(D^2+\lambda\sigma_{cl}^2)\Pi\nonumber\\
&+&\frac{1}{2}\sigma_{cl}D^2\sigma_{cl}+\frac{\lambda}{4}\sigma_{cl}^4
+\frac{e}{2}D^{\alpha}\Pi A_{\alpha}\Sigma
-\frac{e}{2}D^{\alpha}\Sigma A_{\alpha}\Pi
-\frac{e^2}{2}(\Sigma^2+\Pi^2)A^2-e^2\sigma_{cl}\Sigma A^2\\
&+&\frac{\lambda}{4}(\Sigma^4+\Pi^4)
+\frac{\lambda}{2}\Sigma^2\Pi^2+\lambda \sigma_{cl}\Sigma(\Sigma^2+\Pi^2)-eD^{\alpha}\sigma_{cl} \Pi A_{\alpha}
+(\lambda \sigma_{cl}^3+D^2\sigma_{cl})\Sigma \Big{\}}~.\nonumber
\end{eqnarray}

In the quantization process, we may eliminate the mixing between the superfields $A^{\alpha}$ and $\Pi$ that appears in third term of Eq.(\ref{ceq6}), by using an $R_{\xi}$ gauge fixing, $\mathcal{F}_{G}=(D^{\alpha}A_{\alpha}+\alpha e\sigma_{cl}\Pi/2)$. This is done by the inclusion of the gauge fixing and Faddeev-Popov action,
\begin{eqnarray}\label{ceq6a}
S_{GF+FP}&=&\int{d^5z}\Big[\frac{1}{2\alpha}
(D^{\alpha}A_{\alpha}+\alpha\frac{e\sigma_{cl}}{2}\Pi)^2
+\bar{c}D^2c+\frac{\alpha}{4}e^2\sigma_{cl}^2\bar{c}c+\frac{\alpha}{4}{e^2\sigma_{cl}}\bar{c}\Sigma c\Big]~.
\end{eqnarray}

We must observe that even by using an $R_\xi$ gauge a complete elimination of the mixing is not attained; in fact the term $-e D^{\alpha} \sigma_{cl} \Pi A_{\alpha}$ remains. This term will be disregarded in the approximation we will consider below, because it contains supercovariant derivatives of the classical superfield $\sigma_{cl}$.

From the last two terms of Eq.(\ref{ceq6}), we can see that the model can exhibit a nonvanishing tadpole contribution. To avoid problems with the quantization of this theory (e.g. lack of unitarity), we have to impose that the tadpole equation vanish: this condition is usually called the gap equation. At the tree level, the gap equation for $\Sigma$ only allows the trivial solution $\sigma_{cl}=0$. In the sequel, by using the tadpole method, we will calculate the one- and two-loop corrections to the effective superpotential, in an approximation (K\"{a}hlerian) that we will discuss first.

From Eq.(\ref{ceq6}), the zero loop effective action $\Gamma_{cl}$, the superpotential $V_{cl}$, the K\"{a}hlerian superpotential $K_{cl}$ and the scalar potential $U_{cl}$ can be seen to be given by
\begin{eqnarray}\label{ceq6b}
\Gamma[\sigma_{cl}]&=&-\int d^5z~V_{cl}=\int{d^5z}~\left[\frac{1}{2}\sigma_{cl}D^2\sigma_{cl}-K_{cl}(\sigma_{cl}~)\right] \nonumber\\
&=&-\int{d^3x}~U_{cl}=\int{d^3x}~\left[\frac{1}{2}\sigma_2^2-\sigma_2\frac{dK_{cl}}{d\sigma_{cl}}(\sigma_1)~\right] 
\end{eqnarray}
\noindent
where $K_{cl}(\sigma_{cl})=-\dfrac{\lambda}{4} \sigma_{cl}^4$ and $\sigma_2 \dfrac {dK_{cl}}{d\sigma}(\sigma_1)=
-\lambda \sigma_2 \sigma_1^3$. After the elimination of the auxiliary field $\sigma_2$ the scalar potential at classical level is given by:
$U^0_{eff}= \dfrac{\lambda^2}{2}\, \sigma^6_1$.

As we are working in an explicit supersymmetric formulation, the radiative corrections to the action and effective superpotential will be of the form~\cite{poteff3D}
\begin{eqnarray}\label{cweq0b}
\Gamma_{rc}[\sigma_{cl}]=-\int d^5z~V_{rc}
=-\int d^5z~\left[K_{rc}(\sigma_{cl})+~F(D^{\alpha}\sigma_{cl}D_{\alpha}\sigma_{cl},~D^2\sigma_{cl},~\sigma_{cl})\right],
\end{eqnarray}

\noindent
where $K_{rc}(\sigma_{cl})$, a function of $\sigma_{cl}$ but not of its derivatives, stands for the radiative corrections to the K\"{a}hlerian effective superpotential, and $F$ stands for the radiative corrections explicitly involving at least a derivative of the classical field $\sigma_{cl}$. After integrating in $d^2\theta$ we get for $\Gamma_{rc}$ : 
\begin{eqnarray}\label{+1}
\Gamma_{rc}[\sigma_{cl}]
=-\int d^3x\, U_{rc}=-\int d^3x ~\left[~\sigma_2\frac{dK_{rc}}{d\sigma_1}(\sigma_1)+\sigma_2^2 f(\sigma_1,\sigma_2)\right]
\end{eqnarray}

\noindent
where in the second term we made explicit the fact that the contributions coming from the $F$ term start at least
with two powers of $\sigma_2$.

From Eq.(\ref{ceq6b}), Eq.(\ref{cweq0b}) and Eq.(\ref{+1}), we have for the effective superpotential and the 
effective scalar potential:
\begin{eqnarray}\label{cweq0c}
V_{eff}(\sigma_{cl})&=&-\frac{1}{2}\sigma_{cl}D^2\sigma_{cl}+F(D^{\alpha}\sigma_{cl}D_{\alpha}\sigma_{cl},\,D^2\sigma_{cl},\,\sigma_{cl})+K(\sigma_{cl})\nonumber\\
U_{eff}(\sigma_1,\sigma_2)&=&-\frac{1}{2} \sigma_2^2 +\sigma_2^2 f(\sigma_1,\sigma_2) +\sigma_2\frac{dK}{d\sigma_1}(\sigma_1)~,
\end{eqnarray}
 
\noindent
where $K=K_{cl}+K_{rc}$. The vacuum of the model is determined by the equations:
\begin{eqnarray}
0&=&\frac{\partial U_{eff}}{\partial \sigma_1}=\sigma_2 \frac{d^2K}{d \sigma_1^2}(\sigma_1)+ 
\sigma_2^2\frac{\partial f}{\partial \sigma_1}(\sigma_1,\sigma_2),\\
0&=&\frac{\partial U_{eff}}{\partial \sigma_2}=-\sigma_2+\frac{d K}{d\sigma_1}(\sigma_1)
+2 \sigma_2 f(\sigma_1,\sigma_2)+\sigma_2^2 \frac{\partial f}{\partial \sigma_2}(\sigma_2,\sigma_2).
\end{eqnarray}

\noindent
For $\sigma_2=0$ these equations result in $U_{eff}(\sigma_1,0)=0$ at its minimum, signaling a supersymmetric phase, and the condition:
\begin{eqnarray}
0=\frac{d K}{d \sigma_1}(\sigma_1).
\end{eqnarray}

\noindent
So, if a solution $\sigma_1=v$ of this last equation exists, then $U_{eff}(v,0)=0$, and supersymmetry is preserved.
In short: the verification of the preservation of supersymmetry only requires the knowledge of $K(\sigma_{cl})$.
From now on we will restrict to its calculation, instead of the more involved calculation of $V_{eff}(\sigma_{cl})$.
 
To calculate $K(\sigma_{cl})$ it is enough to derive the Feynman rules from Eq.(\ref{ceq6}) and Eq.(\ref{ceq6a}) by
only preserving the dependence in $\sigma$ and dropping dependences in $D_{\alpha}\sigma$ and $D^2\sigma$, 
which also means, to make the D-algebra operations by taking $D_{\alpha}(\sigma X)= \sigma D_{\alpha}X$ and 
$D^2( \sigma X)=\sigma D^2 X$. In this way the free propagators are given by:
\begin{eqnarray}\label{props}
\langle T~\Sigma(k,\theta)\Sigma(-k,\theta')\rangle&=&-i\frac{D^2-M_{\Sigma}}{k^2+M_{\Sigma}^2}\delta^{(2)}(\theta-\theta')~,\nonumber\\
\langle T~\Pi(k,\theta)\Pi(-k,\theta')\rangle&=&-i\frac{D^2-M_{\Pi}}{k^2+M_{\Pi}^2}\delta^{(2)}(\theta-\theta')~,\\
\langle T~A_{\alpha}(k,\theta)A_{\beta}(-k,\theta')\rangle&=&\frac{i}{4}
\Big[\frac{(D^2+M_{A})D^2D_{\beta}D_{\alpha}}{ k^2(k^2+M_{A}^2)}\nonumber\\
&+&\alpha\frac{(D^2-\alpha M_{A})D^2D_{\alpha}D_{\beta}}
{k^2(k^2+\alpha^2M_{A}^2)}\Big]\delta^{(2)}(\theta-\theta')~.\nonumber
\end{eqnarray}

\noindent
It is important to remark that the effective superpotential is a gauge-dependent quantity, as discussed in~\cite{Jackiw:1974cv}. For simplicity, we will work in a supersymmetric Landau gauge, that is $\alpha=0$. With this choice, the ghosts superfields decouple, and we can identify the ``masses''~of the interacting superfields as
\begin{eqnarray}\label{mass}
M_{\Sigma}=3\lambda\sigma_{cl}^2, \quad M_{\Pi}=\lambda\sigma_{cl}^2~, \quad M_{A}=\frac{e^2\sigma_{cl}^2}{4}~.
\end{eqnarray}

\begin{figure}[ht]
 \begin{center}
\includegraphics[]{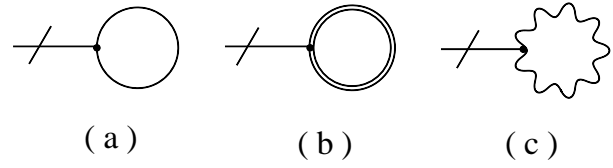}
  \end{center}
\caption{ One-loop contribution to the tadpole equation in the Landau gauge ($\alpha=0$). Solid lines represent the $\Sigma$ propagator, double lines the $\Pi$ propagator, wave lines the gauge superfield propagator and cut solid lines $\Sigma$ superfield.}  \label{gap1l}
\end{figure}

First, let us consider the one-loop corrections to the tadpole equation, which are drawn in Fig.~\ref{gap1l}. The contributions at one-loop order, up to a common 
$\int \!\frac{d^3 p}{(2 \pi)^3} d^2\theta \,  \Sigma(p,\theta)$ factor, can be written as   
\begin{subequations}
\label{ceq6c}
\begin{eqnarray}
&&\Gamma^{(1)}_{1l(a)}=3i\lambda \sigma_{cl}
\int\frac{d^3k}{(2\pi)^3}\frac{1}{k^2+M_{\Sigma}^2}~, \\
&&\Gamma^{(1)}_{1l(b)}=i\lambda\sigma_{cl}~
\int\frac{d^3k}{(2\pi)^3}\frac{1}{k^2+M_{\Pi}^2}~,\\
&&\Gamma^{(1)}_{1l(c)}=i\frac{e^2}{4}\sigma_{cl}~
\int\frac{d^3k}{(2\pi)^3}\frac{1}{k^2+M_{A}^2}~.
\end{eqnarray} 
\end{subequations}

We will perform the integrals in Eq.~(\ref{ceq6c}) by using the regularization by dimensional reduction ~\cite{Siegel:1979wq}. This method of regularization has some advantages: first, one-loop graphs are finite; second, two-loop divergences are a simple pole in $\epsilon=3-D$; and third, starting with a Lagrangian with only massless parameters, no dimensional parameters are generated (the $\mu$ 
parameter introduced by the regularization will only appears in logarithms). This means, for 
example, that no mass term of the form $\bar \Phi \Phi$ (which is not present in the unrenormalized 
Lagrangian) is generated by the radiative corrections. For the tadpoles we get
\begin{eqnarray}\label{ceq6d}
\Gamma^{(1)}_{(0+1)l}=i\frac{\sigma_{cl}^3}{4\pi}\left(4\pi\lambda -10\lambda^2-\frac{e^4}{16}\right)~.
\end{eqnarray} 

\noindent
Setting $\Gamma^{(1)}_{(0+1)l}=0$, we can see that the one-loop correction is not enough to ensure a nontrivial solution to the gap equation, therefore no mass is generated in the first quantum approximation. The same happens in the supersymmetric Maxwell theory~\cite{Burgess:1983nu}, and nonsupersymmetric three-dimensional gauge models~\cite{Tan:1996kz,Tan:1997ew,Dias:2003pw,Alves:1999hw,deAlbuquerque:2000ec}.

Now, we evaluate the two-loop contributions to the tadpole equation, which arise from the diagrams depicted in Fig.~\ref{gap2l}. Some details of this calculations are given in Appendix~\ref{2lcalcs}. The inclusion of such two-loop corrections in the tadpole equation leads to
\begin{eqnarray}\label{ceq8}
\Gamma^{(1)}_{(0+1+2)l}=i~\sigma_{cl}^3\left[b_1+b_2~\ln{\frac{\sigma_{cl}^2}{\mu}}\right]+iB\sigma_{cl}^3~,
\end{eqnarray}

\noindent
where $B$ is the counterterm, to be fixed later, $\mu$ is the mass scale introduced by the regularization, $b_1$ is a function of the coupling constants and $1/\epsilon=1/(3-D)$ ($D$ is the dimension of the spacetime). The quantity $b_2$ is explicitly given by 
\begin{eqnarray}\label{ceq8a}
b_2 &=& -\frac{116~e^6+543~e^4\lambda+432e^2\lambda^2-71552~\lambda^3}{12288\pi^2}\nonumber\\
&\approx& -(10^{-3})~e^6 -(4\times10^{-3})~e^4\lambda -(4\times10^{-3})~e^2\lambda^2 +0.6~\lambda^3~.
\end{eqnarray}

Let us evaluate the K\"{a}hlerian effective superpotential through the tadpole equation Eq.(\ref{ceq11}) using the tadpole method~\cite{Weinberg:1973ua,Miller:1983fe,Miller:1983ri} as in~\cite{Lehum:2008ue}. The K\"{a}hlerian effective superpotential is obtained from the tadpole equation as $K(\sigma_{cl})=i\int{d\sigma}\Gamma^{(1)}$, therefore the two-loop K\"{a}hlerian effective superpotential is given by
\begin{eqnarray}\label{ceq11a}
K(\sigma_{cl})&=&i\int{d\sigma_{cl}}\Gamma^{(1)}_{(0+1+2)l}\nonumber\\
&=&-\frac{b_2}{4}\sigma_{cl}^4\left(\frac{b_1}{b_2}-\frac{1}{2}+\ln{\frac{\sigma_{cl}^2}{\mu}}\right)-\frac{B}{4}\sigma_{cl}^4~.
\end{eqnarray}

\noindent
The normalization of the K\"{a}hlerian effective superpotential, as usual, is defined in terms of the tree-level coupling constant $\lambda$ according to
\begin{eqnarray}
\label{cweq9}
\frac{\lambda}{4}\equiv\frac{1}{4!}\frac{\partial^4 K(\sigma_{cl})}{\partial \sigma_{cl}^4}\Big{|}_{\sigma_{cl}=v}=-\frac{1}{12}\left(3B+3b_1+11b_2+3b_2\ln\frac{v^2}{\mu}\right)~,
\end{eqnarray}

\noindent
where $v$ is the renormalization point; Eq.(\ref{cweq9}) fixes the counterterm $B$,
\begin{eqnarray}
\label{cweq10}
B=-b_1-\lambda-b_2\left(\frac{11}{3}+\ln\frac{v^2}{\mu}\right)~.
\end{eqnarray}

\noindent
By substituting this counterterm in Eq.(\ref{ceq8}), the renormalized K\"{a}hlerian effective superpotential can be cast as
\begin{eqnarray}\label{ceq11}
K(\sigma_{cl})=-\frac{b_2}{4}\sigma_{cl}^4
\ln\left[\frac{\sigma_{cl}^2}{v^2}\exp\left(\frac{\lambda}{b_2}-\frac{25}{6}\right)\right]~.
\end{eqnarray}

The minimum of the renormalized K\"{a}hlerian effective superpotential is at $\sigma_{cl}=v$ that satisfies,
\begin{eqnarray}\label{ceq11aa}
\frac{\partial K(\sigma_{cl})}{\partial\sigma_{cl}}\Big{|}_{\sigma_{cl}=v}=0~,
\end{eqnarray}

\noindent
and a nontrivial minimum, with $v \neq 0$, requires the following constraint on the coupling constants,
\begin{eqnarray}\label{ceq11ab}
\lambda=\dfrac{11}{3}b_2\approx -(4\times10^{-3})~e^6 -(16\times 10^{-3})~e^4\lambda -(13\times 10^{-3})~e^2\lambda^2 +2\lambda^3.
\end{eqnarray}

\noindent
The compatibility of this relation with the assumptions of perturbation theory is the key of Coleman-Weinberg mechanism. From Eq.~(\ref{ceq11ab}), one can see that $\lambda$ must be of order of $(4\times10^{-3})e^6+\mathcal{O}(e^{10})$, thus for small $e$ we are safely within the regime of validity of the perturbative expansion. In a model with vanishing $e$, however, the dynamical gauge symmetry breaking is incompatible with perturbation theory~\cite{Coleman:1973jx,Tan:1996kz,Tan:1997ew,Dias:2003pw,Alves:1999hw,deAlbuquerque:2000ec}, because for $e=0$ the Eq.(\ref{ceq11ab}) implies $\lambda\sim1$.

The mass term of matter superfield $\sigma_{cl}$ is obtained as the second derivative of the renormalized K\"{a}hlerian effective superpotential at its minimum. Using Eq.(\ref{ceq11ab}), we obtain
\begin{eqnarray}\label{ceq11ac}
M_{\Sigma}&=&\frac{d^2K(\sigma_{cl})}{d\sigma_{cl}^2}\Big|_{\sigma_{cl}=v}\approx{(2\times 10^{-3})e^6}v^2~,
\end{eqnarray}

\noindent
where the mass of the gauge superfield is given as 
\begin{eqnarray}\label{ceq12}
{M_A}=\frac{e^2}{12\lambda}~{M_{\Sigma}}\approx -250~e^{-4}~.
\end{eqnarray}

\section{Extended SUSY model}\label{N=2}

A supersymmetric extension of the model defined in Eq.(\ref{ceq1}) to ${\cal{N}}=2$ can be obtained by identifying the coupling constants as $\lambda=-e^2/8$, plus some others generalizations that can be seen in~\cite{Lee:1990it}. {In particular, the ${\cal{N}}=2$ supersymmetric Chern-Simons action written in terms of ${\cal{N}}=1$ superfields was first obtained by Siegel~\cite{Siegel:1979fr}}. Evaluating the quantum corrections discussed in the previous section in this case, the K\"{a}hlerian effective superpotential can be expressed by
\begin{eqnarray}\label{ceq14}
K(\sigma_{cl})=\frac{c_2}{4}e^6\sigma^4\log\Big{\{}\frac{\sigma^2}{v^2}\exp{\left(\frac{1}{c_2e^4}-\frac{25}{6}\right)}\Big{\}},
\end{eqnarray}

\noindent
where $c_2=\dfrac{519}{32768\pi^2}\approx (1.6\times 10^{-3})$. 

Now, studying a possible minimum of such a K\"{a}hlerian effective superpotential for $v \neq 0$, we found that $e^2\approx (4.7\times 10^{-2})~e^6$, which implies $e\sim2$. Thus, as in other four and three-dimensional scalar models~\cite{Coleman:1973jx,Tan:1996kz,Tan:1997ew,Dias:2003pw,Alves:1999hw,deAlbuquerque:2000ec}, in this case the Coleman-Weinberg effect is not feasible, because for ${\mathcal{N}}=2$ the model presents only one coupling constant into play. Our results are in agreement with previous results obtained by Gaiotto and Yin~\cite{Gaiotto:2007qi} {and Buchbinder et al.~\cite{Buchbinder:2010em,Buchbinder:2009dc}}, where it was shown that the ${\cal{N}}=2,3$ SCSM does not exhibit spontaneous breaking of superconformal invariance.

\section{Concluding remarks}\label{final}

In this paper we studied the mechanism of dynamical breaking of gauge symmetry in a three-dimensional supersymmetric model with classical superconformal invariance. Since we were working in an explicitly supersymmetric formalism, no breakdown of supersymmetry was detected, but we show how the two-loop quantum corrections dynamically break the gauge and superconformal invariances of the model, generating masses to the gauge and matter superfields. We also show that, in the ${\cal{N}}=2$ extension of the model studied by us, the Coleman-Weinberg mechanism of dynamical symmetry breaking is incompatible with perturbation theory, in agreement with the results obtained by Gaiotto and Yin~\cite{Gaiotto:2007qi} and Buchbinder et al.~\cite{Buchbinder:2010em}. 

One natural extension of our work would be to consider {the possibility of supersymmetry breaking in the spirit of what was done for the Wess-Zumino model in~\cite{Amariti:2009kb}, or through the component formulation as in~\cite{Lehum:2008vn}.} Another possibility is to study the effective superpotential for a noncommutative extension of the present model, as done in~\cite{poteff3D}.

\vspace{1cm}
{\bf Acknowledgments.} This work was partially supported by the Brazilian agencies Conselho 
Nacional de Desenvolvimento Cient\'{\i}fico e Tecnol\'{o}gico (CNPq), Coordena\c c\~ao de Aperfei\c coamento de Pessoal de N\'\i vel Superior (CAPES: AUX-PE-PROCAD 579/2008) and Funda\c{c}\~{a}o de Amparo \`{a} Pesquisa do Estado de S\~{a}o Paulo (FAPESP). A. Yu. P. has been supported by the CNPq Project No. 303461-2009/8.

\appendix

\section{The expansions of superfields}\label{sup}

A complex superfield can be written as the sum of real and imaginary parts as
\begin{eqnarray}\label{sup1}
\Phi(x,\theta)&=&\frac{1}{\sqrt{2}}\left[\Sigma(x,\theta)+i\Pi(x,\theta)\right]~,\nonumber\\
\bar\Phi(x,\theta)&=&\frac{1}{\sqrt{2}}\left[\Sigma(x,\theta)-i\Pi(x,\theta)\right]~,
\end{eqnarray}

\noindent
where the real scalar superfields can be expanded in terms of component fields as follows
\begin{eqnarray}\label{sup2}
\Sigma(x,\theta)&=&\sigma(x)+\theta^{\alpha}\psi_{\alpha}(x)-\theta^2F(x)~,\nonumber\\
\Pi(x,\theta)&=&\pi(x)+\theta^{\alpha}\xi_{\alpha}(x)-\theta^2G(x)~.
\end{eqnarray}

The spinorial superfield (gauge superfield), possess the following $\theta$ expansion
\begin{eqnarray}\label{sup3}
A_{\alpha}(x,\theta)=\chi_{\alpha}(x)+\theta^{\beta}\left[C_{\alpha\beta}B(x)-iV_{\alpha\beta}(x)\right]
-\theta^2~\tilde{\lambda}_\alpha(x)~,
\end{eqnarray}

\noindent
and the associated field strength, 
\begin{eqnarray}\label{sup4}
W_{\alpha}=\frac{1}{2}D^{\beta}D_{\alpha}A_{\beta}~,
\end{eqnarray}

\noindent
satisfies the Bianchi identity $D^{\alpha}W_{\alpha}=0$. The component expansion of this field strength reads
\begin{eqnarray}\label{sup5}
W_{\alpha}=\frac{1}{2}
\left({\partial_{\alpha}}^{\beta}\chi_{\beta}+\tilde{\lambda}_{\alpha}\right)
+\theta^{\beta}f_{\beta\alpha}
-\theta^2\frac{1}{2}\left(i{\partial_{\alpha}}^{\beta}\tilde{\lambda}_{\beta}+\Box\chi_{\alpha}\right)~,
\end{eqnarray} 

\noindent
where $f_{\alpha\beta}=(\partial_{\alpha\lambda}{V^{\lambda}}_{\beta}+\partial_{\beta\lambda}{V^{\lambda}}_{\alpha})$ 
is the 2-spinorial form of the Maxwell field strength 
$F_{MN}=(\partial_{M}A_{N}-\partial_{N}A_{M})$, $M$ and $N$ are the usual indices of the spacetime, which assume the values $0$, $1$ and $2$.

\section{Two-loop calculations}\label{2lcalcs}

\begin{figure}[ht]
 \begin{center}
\includegraphics[]{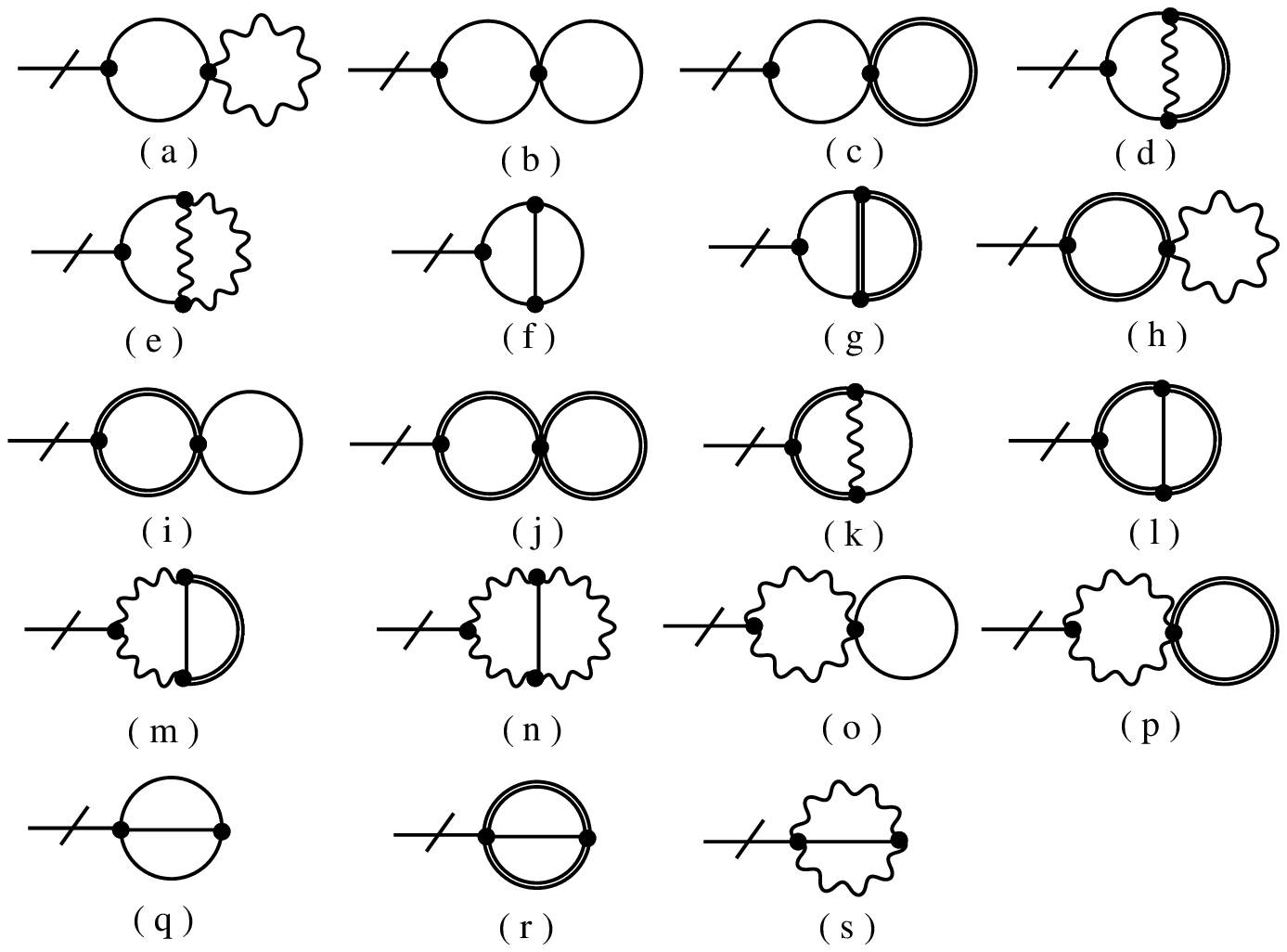}
  \end{center}
\caption{ Diagrams that contribute to the tadpole equation at two-loop approximation in the supersymmetric Landau gauge.}\label{gap2l}
\end{figure}

At two-loop, the contributions to the gap equation, associated with the diagrams depicted in Fig.~\ref{gap2l}, are given by
\begin{eqnarray}\label{ceq7}
\Gamma^{(1)}_{2l(a)}&=&i\frac{3e^2}{2}\lambda\sigma_{cl}~
\int\frac{d^3k}{(2\pi)^3}\frac{d^3q}{(2\pi)^3}\frac{M_{\Sigma}}{(k^2+M_{\Sigma}^2)^2(q^2+M_A^2)}~,
\end{eqnarray}
\begin{eqnarray}\label{ceq7a}
\Gamma^{(1)}_{2l(b)}&=&-{18i}\lambda^2\sigma_{cl}~
\int\frac{d^3k}{(2\pi)^3}\frac{d^3q}{(2\pi)^3}\frac{M_{\Sigma}}{(k^2+M_{\Sigma}^2)^2(q^2+M_{\Sigma}^2)}~,
\end{eqnarray}
\begin{eqnarray}\label{ceq7b}
\Gamma^{(1)}_{2l(c)}&=&-{3i}\lambda^2\sigma_{cl}~
\int\frac{d^3k}{(2\pi)^3}\frac{d^3q}{(2\pi)^3}\frac{M_{\Sigma}}{(k^2+M_{\Sigma}^2)^2(q^2+M_{\Pi}^2)}~,
\end{eqnarray}
\begin{eqnarray}\label{ceq7c}
\Gamma^{(1)}_{2l(d)}&=&-i\frac{3}{4}\lambda~e^2\sigma_{cl}~
\int\frac{d^3k}{(2\pi)^3}\frac{d^3q}{(2\pi)^3}
\frac{1}{(k^2+M_{\Sigma}^2)^2[(k+q)^2+M_{\Pi}^2]q^2(q^2+M_{A}^2)}\nonumber\\
&&\times\Big[M_{\Sigma}^2M_A~k\cdot q-2M_A M_{\Sigma} M_{\Pi}~k\cdot q 
- M_A~k^2~k\cdot q -(M_A+M_{\Pi})M_{\Sigma}^2~q^2\nonumber\\
&&-(M_A+M_{\Pi}-2M_{\Sigma})~q^2~k^2+2M_{\Sigma}~q^2~k\cdot q\Big]~,
\end{eqnarray}
\begin{eqnarray}\label{ceq7d}
\Gamma^{(1)}_{2l(e)}&=&\dfrac{3i}{16}\lambda e^4\sigma_{cl}^3~
\int\frac{d^3k}{(2\pi)^3}\frac{d^3q}{(2\pi)^3}
\frac{1}{k^2(k^2+M_{A}^2)[(k+q)^2+M_{\Sigma}^2]^2q^2(q^2+M_{A}^2)}\nonumber\\
&\times&\Big[M_{\Sigma}^2M_A^2~(k\cdot q) - M_A^2~(k+q)^2~(k\cdot q)
-2 M_A M_{\Sigma}~(k\cdot q)~(k^2+q^2) \nonumber\\
&&+4 M_A M_{\Sigma}~k^2~q^2+M_{\Sigma}^2~k^2~q^2-k^2~q^2~(k+q)^2\Big]~,
\end{eqnarray}
\begin{eqnarray}\label{ceq7e}
\Gamma^{(1)}_{2l(f)}&=&-18i\lambda^3\sigma_{cl}^3~
\int\frac{d^3k}{(2\pi)^3}\frac{d^3q}{(2\pi)^3}
\frac{5M_{\Sigma}^2-k^2}{(k^2+M_{\Sigma}^2)^2[(k+q)^2+M_{\Sigma}^2](q^2+M_{\Sigma}^2)}~,
\end{eqnarray}
\begin{eqnarray}\label{ceq7f}
\Gamma^{(1)}_{2l(g)}&=&-6i\lambda^3\sigma_{cl}^3~
\int\frac{d^3k}{(2\pi)^3}\frac{d^3q}{(2\pi)^3}
\frac{M_{\Sigma}^2+4M_{\Sigma}M_{\Pi}-k^2}{(k^2+M_{\Sigma}^2)^2[(k+q)^2+M_{\Pi}^2](q^2+M_{\Pi}^2)}~,
\end{eqnarray}
\begin{eqnarray}\label{ceq7g}
\Gamma^{(1)}_{2l(h)}&=&-i\frac{e^2}{2}\lambda \sigma_{cl}~
\int\frac{d^3k}{(2\pi)^3}\frac{d^3q}{(2\pi)^3}\frac{M_{\Pi}}{(k^2+M_{\Pi}^2)^2(q^2+M_A^2)}~,
\end{eqnarray}
\begin{eqnarray}\label{ceq7h}
\Gamma^{(1)}_{2l(i)}&=&-2i\lambda^2\sigma_{cl}~
\int\frac{d^3k}{(2\pi)^3}\frac{d^3q}{(2\pi)^3}\frac{M_{\Pi}}{(k^2+M_{\Pi}^2)^2(q^2+M_{\Sigma}^2)}~,
\end{eqnarray}
\begin{eqnarray}\label{ceq7i}
\Gamma^{(1)}_{2l(j)}&=&-6i\lambda^2\sigma_{cl}~
\int\frac{d^3k}{(2\pi)^3}\frac{d^3q}{(2\pi)^3}\frac{M_{\Pi}}{(k^2+M_{\Pi}^2)^2(q^2+M_{\Pi}^2)}~,
\end{eqnarray}
\begin{eqnarray}\label{ceq7j}
\Gamma^{(1)}_{2l(k)}&=&\frac{i}{4}\lambda e^2\sigma_{cl}~
\int\frac{d^3k}{(2\pi)^3}\frac{d^3q}{(2\pi)^3}
\frac{1}{(k^2+M_{\Pi}^2)^2[(k+q)^2+M_{\Sigma}^2]q^2(q^2+M_{A}^2)}\nonumber\\
&\times&\Big[M_AM_{\Pi}(2M_{\Sigma}-M_{\Pi})~k\cdot q
- M_A~(k\cdot q)~k^2 - M_{\Pi}^2(M_A+M_{\Sigma})~q^2\nonumber\\ 
&&-2M_{\Pi}~q^2~k\cdot q+(M_A+M_{\Sigma}-2M_{\Pi})~k^2~q^2\Big]~,
\end{eqnarray}
\begin{eqnarray}\label{ceq7k}
\Gamma^{(1)}_{2l(l)}&=&-\frac{4i}{3}\lambda^3\sigma_{cl}^3~
\int\frac{d^3k}{(2\pi)^3}\frac{d^3q}{(2\pi)^3}\frac{3M_{\Pi}+2M_{\Pi}M_{\Sigma}-k^2}
{(k^2+M_{\Pi}^2)^2(q^2+M_{\Pi})[(k+q)^2+M_{\Sigma}]}~,
\end{eqnarray}
\begin{eqnarray}\label{ceq7l}
\Gamma^{(1)}_{2l(m)}&=&i\frac{e^4}{8}\sigma_{cl}~
\int\frac{d^3k}{(2\pi)^3}\frac{d^3q}{(2\pi)^3}
\frac{1}{k^2(k^2+M_{A}^2)^2[(k+q)^2+M_{\Pi}^2](q^2+M_{\Sigma}^2)}\nonumber\\
&\times&\Big[M_A^2(M_{\Pi}-M_{\Sigma})~k\cdot q
-M_{\Sigma}M_A(M_A+2M_{\Pi})~k^2\nonumber\\
&&+(M_{\Sigma}-2M_A-M_{\Pi})~(k\cdot q)~k^2 +M_{\Sigma}~(k^2)^2 
-2M_{A}~k^2~q^2\Big]~,
\end{eqnarray}
\begin{eqnarray}\label{ceq7m}
\Gamma^{(1)}_{2l(n)}&=&-\frac{i}{96}e^6\sigma_{cl}^3~
\int\frac{d^3k}{(2\pi)^3}\frac{d^3q}{(2\pi)^3}
\frac{1}{k^2(k^2+M_{A}^2)^2[(k+q)^2+M_{\Sigma}^2]q^2(q^2+M_{A}^2)}\nonumber\\
&\times&\Big[M_A^3M_{\Sigma}~k\cdot q-(k^2)^2~q^2
-M_A(M_{\Sigma}+M_A)~k\cdot q~k^2\nonumber\\
&& -M_A^2~k\cdot q~q^2-M_A(3M_A+2M_{\Sigma})~k^2~q^2+(k\cdot q)~k^2~q^2
\Big]~,
\end{eqnarray}
\begin{eqnarray}\label{ceq7n}
\Gamma^{(1)}_{2l(o)}&=&i\frac{e^4}{16}\sigma_{cl}~
\int\frac{d^3k}{(2\pi)^3}\frac{d^3q}{(2\pi)^3}\frac{M_{A}}{(k^2+M_{A}^2)^2(q^2+M_{\Sigma}^2)}~,
\end{eqnarray}
\begin{eqnarray}\label{ceq7o}
\Gamma^{(1)}_{2l(p)}&=&i\frac{e^4}{16}\sigma_{cl}~
\int\frac{d^3k}{(2\pi)^3}\frac{d^3q}{(2\pi)^3}\frac{M_{A}}{(k^2+M_{A}^2)^2(q^2+M_{\Pi}^2)}~,
\end{eqnarray}
\begin{eqnarray}\label{ceq7q}
\Gamma^{(1)}_{2l(q)}&=&-9i\lambda^2\sigma_{cl}~
\int\frac{d^3k}{(2\pi)^3}\frac{d^3q}{(2\pi)^3}
\frac{M_{\Sigma}}{(k^2+M_{\Sigma}^2)(q^2+M_{\Sigma}^2)[(k-q)^2+M_{\Sigma}^2]}~,
\end{eqnarray}
\begin{eqnarray}\label{ceq7r}
\Gamma^{(1)}_{2l(r)}&=&-\frac{4i}{3}\lambda^2\sigma_{cl}~
\int\frac{d^3k}{(2\pi)^3}\frac{d^3q}{(2\pi)^3}
\frac{M_{\Sigma}+2M_{\Pi}}{(k^2+M_{\Sigma}^2)(q^2+M_{\Pi}^2)[(k-q)^2+M_{\Pi}^2]}~,
\end{eqnarray}
\begin{eqnarray}\label{ceq7s}
\Gamma^{(1)}_{2l(s)}&=&-i\frac{e^4}{16}\sigma_{cl}~
\int\frac{d^3k}{(2\pi)^3}\frac{d^3q}{(2\pi)^3}
 \frac{M_{A}^2M_{\Sigma}~k\cdot q -M_A(k^2+q^2)~k\cdot q-(2M_A+M_{\Sigma})~k^2~q^2}
{k^2(k^2+M_{A}^2)q^2(q^2+M_{A}^2)[(q-k)^2+M_{\Sigma}^2]}~,
\end{eqnarray}

\noindent
where the D-algebra manipulations on the two-loop supergraphs were performed with the help of  SusyMath~\cite{Ferrari:2007sc}, a MATHEMATICA$^{\copyright}$ package for supergraph calculations.

Performing the integrals in Eqs.~(\ref{ceq7}-\ref{ceq7s}) with the help of formulas~\cite{Tan:1996kz,Tan:1997ew,Dias:2003pw}, adopting the regularization by dimensional reduction~ \cite{Siegel:1979wq}, and adding up the classical, one and two-loop contributions, we obtain the following one point function,
\begin{eqnarray}\label{ceq82l}
\Gamma^{(1)}_{(0+1+2)l}=i~\sigma_{cl}^3\left[b_1+b_2~\ln{\frac{\sigma_{cl}^2}{\mu}}\right]~,
\end{eqnarray}

\noindent
where $\mu$ is a mass scale introduced by the regularization, 
\begin{eqnarray}\label{eqb2}
b_2&=&-\dfrac{116~e^6+543~e^4\lambda+432e^2\lambda^2-71552~\lambda^3}{12288\pi^2} \nonumber\\
&\approx &-(10^{-3})~e^6 -(4\times10^{-3})~e^4\lambda -(4\times10^{-3})~e^2\lambda^2 +0.6~\lambda^3,
\end{eqnarray}

\noindent and $b_1$ is some function of coupling constants and $1/\epsilon=1/(3-D)$ ($D$ is the dimension of spacetime).

%\newpage
\vspace{1cm}

\begin{thebibliography}{99}

\bibitem{Coleman:1973jx}
  S.~R.~Coleman and E.~Weinberg,
  Phys.\ Rev.\  D {\bf 7}, 1888 (1973).

\bibitem{Tan:1996kz}
  P.~N.~Tan, B.~Tekin and Y.~Hosotani,
  Phys.\ Lett.\  B {\bf 388}, 611 (1996).

\bibitem{Tan:1997ew}
  P.~N.~Tan, B.~Tekin and Y.~Hosotani,
  Nucl.\ Phys.\  B {\bf 502}, 483 (1997).

\bibitem{Dias:2003pw}
  A.~G.~Dias, M.~Gomes and A.~J.~da Silva,
  Phys.\ Rev.\  D {\bf 69}, 065011 (2004).

\bibitem{Alves:1999hw}
  V.~S.~Alves, M.~Gomes, S.~L.~V.~Pinheiro and A.~J.~da Silva,
  Phys.\ Rev.\  D {\bf 61}, 065003 (2000).

\bibitem{deAlbuquerque:2000ec}
  L.~C.~de Albuquerque, M.~Gomes and A.~J.~da Silva,
  Phys.\ Rev.\  D {\bf 62}, 085005 (2000).
  
  
\bibitem{Deser:1981wh}
  S.~Deser, R.~Jackiw and S.~Templeton,
  Ann. Phys. (N.Y.){\bf 140}, 372 (1982); {\bf 185}, 406(E) (1988); {\bf 281}, 409(E) (2000).

\bibitem{Schonfeld:1981}
J. A. Schonfeld, Nuclear Phys. B {\bf 185}, 157 (1981).

\bibitem{prange}
``The Quantum Hall Effect'', 
Graduate Texts in Contemporary Physics, 
R.E. Prange and S.M. Girvin eds, Springer-Verlag, Berlin, 1990.

\bibitem{Schwarz:2004yj}
  J.~H.~Schwarz,
  JHEP {\bf 0411}, 078 (2004).

\bibitem{Bandres:2008vf}
  M.~A.~Bandres, A.~E.~Lipstein and J.~H.~Schwarz,
  JHEP {\bf 0805}, 025 (2008).

\bibitem{Aharony:2008ug}
  O.~Aharony, O.~Bergman, D.~L.~Jafferis and J.~Maldacena,
  JHEP {\bf 0810}, 091 (2008).

\bibitem{Bagger:2006sk}
  J.~Bagger and N.~Lambert,
  Phys.\ Rev.\  D {\bf 75}, 045020 (2007).

\bibitem{Bagger:2007jr}
  J.~Bagger~and~N.~Lambert,
  Phys.\ Rev.\  D {\bf 77}, 065008 (2008).

\bibitem{Gustavsson:2007vu}
  A.~Gustavsson,
  Nucl.\ Phys.\  B {\bf 811}, 66 (2009).

\bibitem{Miskovic:2009dd}
  O.~Miskovic and J.~Zanelli,
  Phys.\ Rev.\  D {\bf 80}, 044003 (2009).

\bibitem{Burgess:1983nu}
  C.~P.~Burgess,
  Nucl.\ Phys.\  B {\bf 216}, 459 (1983).

\bibitem{Lehum:2008vn}
  A.~C.~Lehum,
  Phys.\ Rev.\  D {\bf 77}, 067701 (2008).

\bibitem{Lehum:2008ue}
  A.~C.~Lehum,
  Phys.\ Rev.\  D {\bf 79}, 025005 (2009).

\bibitem{Gaiotto:2007qi}
  D.~Gaiotto and X.~Yin,
  JHEP {\bf 0708}, 056 (2007).

\bibitem{Buchbinder:2010em}
  I.~L.~Buchbinder, N.~G.~Pletnev and I.~B.~Samsonov,
  ``Effective action of three-dimensional extended supersymmetric matter on
  gauge superfield background,''
  arXiv:1003.4806 [hep-th].

\bibitem{Lee:1990it}
  C.~K.~Lee, K.~M.~Lee and E.~J.~Weinberg,
  Phys.\ Lett.\  B {\bf 243}, 105 (1990).

\bibitem{Gates:1991qn}
  S.~J.~J.~Gates and H.~Nishino,
  Phys.\ Lett.\  B {\bf 281}, 72 (1992).

\bibitem{Nishino:1991sr}
  H.~Nishino and S.~J.~J.~Gates,
  Int.\ J.\ Mod.\ Phys.\  A {\bf 8}, 3371 (1993).

\bibitem{Gates:1983nr}
  S.~J.~Gates, M.~T.~Grisaru, M.~Rocek and W.~Siegel,
  Front.\ Phys.\  {\bf 58}, 1 (1983).

\bibitem{Lehum:2007nf}
  A.~C.~Lehum, A.~F.~Ferrari, M.~Gomes and A.~J.~da Silva,
  Phys.\ Rev.\  D {\bf 76}, 105021 (2007).

\bibitem{poteff3D}
  A.~F.~Ferrari, M.~Gomes, A.~C.~Lehum, J.~R.~Nascimento, A.~Y.~Petrov, E.~O.~Silva and A.~J.~da Silva,
  Phys.\ Lett.\  B {\bf 678}, 500 (2009).

\bibitem{Jackiw:1974cv}
  R.~Jackiw,
  Phys.\ Rev.\  D {\bf 9}, 1686 (1974).

\bibitem{Siegel:1979wq}
  W.~Siegel,
  Phys.\ Lett.\  B {\bf 84}, 193 (1979).

\bibitem{Weinberg:1973ua}
  S.~Weinberg, 
  Phys.\ Rev.\  D {\bf 7}, 2887 (1973).

\bibitem{Miller:1983fe}
  R.~D.~C.~Miller,
  Nucl.\ Phys.\  B {\bf 228}, 316 (1983).

\bibitem{Miller:1983ri}
  R.~D.~C.~Miller,
  Nucl.\ Phys.\  B {\bf 229}, 189 (1983).

\bibitem{Siegel:1979fr}
  W.~Siegel,
  Nucl.\ Phys.\  B {\bf 156}, 135 (1979).

\bibitem{Buchbinder:2009dc}
  I.~L.~Buchbinder, E.~A.~Ivanov, O.~Lechtenfeld, N.~G.~Pletnev, I.~B.~Samsonov and B.~M.~Zupnik,
  JHEP {\bf 0910}, 075 (2009).

\bibitem{Amariti:2009kb}
  A.~Amariti and M.~Siani,
  JHEP {\bf 0908}, 055 (2009)  
  
\bibitem{Ferrari:2007sc}
  A.~F.~Ferrari,
  Comput.\ Phys.\ Commun.\  {\bf 176}, 334 (2007).
  

\end{thebibliography}
\end{document}